\documentclass[12pt]{iopart}
\usepackage{graphicx}

\begin{document}

\title[Spatial fluctuations in transient creep deformation]
{Spatial fluctuations in transient creep deformation}

\author{Lasse Laurson$^{1,2}$, Jari Rosti$^1$, Juha Koivisto$^1$, 
Amandine Miksic$^1$, and Mikko J. Alava$^1$}

\address{$ˆ1$ Department of Applied Physics, Aalto University, 
PO Box 14100, Aalto 00076, Finland}
\address{{$^2$ISI Foundation, Viale S. Severo 65, 10133 Torino, 
Italy}}

\ead{lasse.laurson@aalto.fi}

\begin{abstract}
We study the spatial fluctuations of transient creep
deformation of materials as a function of time, both by Digital 
Image Correlation (DIC) measurements of paper samples and by numerical 
simulations of a crystal plasticity or discrete dislocation dynamics model. 
This model has a jamming or yielding phase transition, around which 
power-law or Andrade creep is found. During primary
creep, the relative strength of the strain rate fluctuations
increases with time in both cases - the spatially averaged creep rate
obeys the Andrade law $\epsilon_t \sim t^{-0.7}$, while the time
dependence of the spatial fluctuations of the local creep rates is
given by $\Delta \epsilon_t \sim t^{-0.5}$. A similar scaling for
the fluctuations is found in the logarithmic creep regime that is
typically observed for lower applied stresses. We review
briefly some classical theories of Andrade creep from the point of
view of such spatial fluctuations. We consider these phenomenological,
time-dependent creep laws in terms of a description
based on a non-equilibrium phase transition separating evolving and
frozen states of the system when the externally applied load is varied.
Such an interpretation is discussed further by the
data collapse of the local deformations in the spirit of absorbing
state/depinning phase transitions, as well as deformation-deformation
correlations and the width of the cumulative strain distributions.
The results are also compared with the order parameter fluctuations 
observed close to the depinning transition of the 2$d$ Linear Interface
Model or the quenched Edwards-Wilkinson equation. 
\end{abstract}
\maketitle

\section{Introduction}
Recent research has highlighted the importance of fluctuations and
heterogeneous response in the behavior of materials. The central
issue is that the "rheological" laws that describe what happens to a
sample under loading are coarse-grained from the microscopic
behavior. From this viewpoint, it is then imperative to understand
what are the origins of various phenomena in this field, and to
grasp the consequences for e.g. materials science. The main emphasis
in what follows starts from the phenomenological primary creep law,
which states that the deformation rate decays in time as a power
law, as $\frac{d \epsilon(t)}{dt} \equiv \epsilon_t \sim
t^{-\theta}$. The particular value $2/3$ is coined as Andrade's
creep, originating from 1910  and its theoretical roots are still under a
debate \cite{andrade,main,louchet}.

Crystalline materials deform plastically via the motion of
dislocations. Some time ago, it was shown that two-dimensional
dislocation assemblies exhibit a \emph{yielding
transition} \cite{miguel2}. This is a non-equilibrium phase transition 
separating a state which is asymptotically quiescent from an active (yielding)
one as the control parameter, external stress, is varied. The order
parameter - the strain rate - seems to exhibit a second order phase
transition at the critical yield stress $\sigma_c$. Since it was
shown that such simple materials science/physics models exhibit
critical phenomena the research related to the statistical mechanics
of crystal plasticity has exploded to various directions including
coarse-grained models \cite{moretti}, studies of the properties of the transition 
\cite{lasse}, experimental signatures such as crackling noise as acoustic emission
\cite{miguel,weiss} and in stress-strain curves \cite{dimiduk}, characterization
of the deformation structures \cite{jakobsen}, surface patterning 
\cite{zaiser_morphology}, and three-dimensional studies \cite{csikor}.

However, e.g. Andrade creep is a concept that has been demonstrated
in non-crystalline materials from rocks to composites \cite{nabarro}. 
To outline some particular cases, the deformation of amorphous metallic 
glasses has recently become an active field of its own, and there as well
collective phenomena are now being discussed and studied \cite{barrat,maloney,
goyon}. In that case, and in the physics of the jamming of granular assemblies 
\cite{dauchot,liu,keys} the origins of the intermittent deformation are 
being studied with concepts that are focused upon such materials, such as 
"shear transformation zones" or the "cage effect" \cite{berthier}. Whatever 
the microscopic physics, it is an important question of what kind of 
signatures can be actually found in experiments on deformation on the
coarse-grained scale. In this work, we tackle this issue by studying
the spatiotemporal characteristics of creep deformation, that is to
say the response of material samples to a constant load. As noted
above, there are indications that collective phenomena might be of
decisive importance. A short account of the results has been
published recently in \cite{main}. 

The questions that are of primary interest are: i) what kind of
fluctuations can be seen in the creep of experimental samples and
idealized dislocation assemblies, by computer simulation? ii) do
these show universality beyond the models that would be appropriate
for the material at hand? iii) what kind of correlations and
fluctuations ensue, for both the "order parameter" (creep rate, for
the Andrade creep at least) and the integrated order parameter, i.e.
the total creep strain? The outcomes for such questions present a
challenge for the statistical physics of materials science. There
are a few theories relevant for the creep deformation of materials.
For the particular case of dislocation systems, it appears that
modified versions of depinning (DP) transitions (of elastic
manifolds in random media) or absorbing state (AS) phase transitions
might be in order \cite{moretti}. The fundamental problem is how to 
coarse-grain such a theory from assemblies of individual topological 
defects \cite{GRO-97};
however the DP/AS paradigm offers suggestions about what to look for
in the empirical data whether from simulations of a model that is
supposed to adhere to such a phase transition picture, or from the
experiments.

In what follows, simple paper samples are found to exhibit a primary
creep regime characterized by the power law decay of the average
creep strain rate, $\epsilon_t \sim t^{-0.7}$, analogously to the
Andrade law valid for many materials ranging from soft metals to
ordinary paper \cite{andrade, coffin}. The spatial variations of the
strain rate are studied by the digital image correlation
method, and are found to be characterized by a power law time
dependence of the standard deviation of the local strain rates,
$\Delta \epsilon_t \sim t^{-0.5}$, during the Andrade's creep. Thus,
the magnitude of the fluctuations of the strain rate decays slower
with time than the average creep rate, implying that fluctuations
become more important in relative terms with time. This feature
persists in the logarithmic creep. Similar behavior is observed in a
discrete dislocation dynamics or crystal plasticity model, which is
known to exhibit a non-equilibrium jamming/yielding phase transition
at a critical value of the applied external shear stress. These
results are further compared with simulations of the two-dimensional
Linear Interface
Model (LIM)/quenched Edwards-Wilkinson (qEW) equation close to the depinning
transition. Such a simple 2$d$ model provides a convenient ``benchmark''
system with a simple AS/DP transition where the relevant phenomena
can be studied in a transparent manner.

The paper is organized as follows: In the next Section we discuss
details of the numerical simulations of the discrete dislocation dynamics
model, as well as the experimental setup and measurement methods of the
paper deformation experiments.
In Section \ref{sec:results} we present the results of both the
numerical simulations, including also simulations of the LIM/qEW case,
and experiments. These are then discussed both from the point of
view of classical theories of Andrade creep in dislocation systems,
as well as by considering a picture emerging from interpreting the
creep deformation as a process occurring close to a non-equilibrium
second order phase transition. Section \ref{sec:concl} finishes 
the paper with conclusions.

\section{Details of simulations and experiments}

\subsection{Dislocation model simulations}

The discrete dislocation dynamics (DDD) simulations are performed by means
of a two-dimensional model of shear deformation with point-like edge
dislocations (See \cite{miguel2} or e.g.~\cite{giessen}) .
These point dislocations glide under the influence of the local
Peach-Koehler forces or shear stresses acting on them. These are
superpositions of the applied external shear stress $\sigma$ and the
internal stresses $\sigma_s$ due to the long range anisotropic
stress fields
\begin{equation}
\sigma_s ({\bf r}) = Dbx\frac{(x^2-y^2)}{(x^2+y^2)^2}
\end{equation}
of all the other dislocations in the system. Here
$D=\mu/2\pi(1-\nu)$, with $\mu$ the shear modulus and $\nu$ the
Poisson ratio of the material. $b$ is the magnitude of the Burgers
vector of the dislocations. Such a system can also be thought to
represent a two-dimensional cross section of a three-dimensional
assembly of straight parallel edge dislocations. Only a single slip
system is considered (i.e. dislocation glide is allowed in the
$x$-direction only within the $xy$ plane of the system), and no
dislocation climb is taken into account. The equations of motion are
taken to be overdamped, such that
\begin{equation}
\frac{\chi_d^{-1}v_n}{b} = s_n b \left[ \sum_{m\neq n} s_m
\sigma_s({\bf r}_{nm}) + \sigma \right],
\end{equation}
where $v_n$ is the velocity of the $n$th dislocation, $\chi_d$ refers
to the dislocation mobility, $s_n$ is the sign of the Burgers vector of
the $n$th dislocation, and $\sigma$ is the external shear stress.

\begin{figure}[t!]
\begin{center}
\includegraphics[
width=8cm,clip]{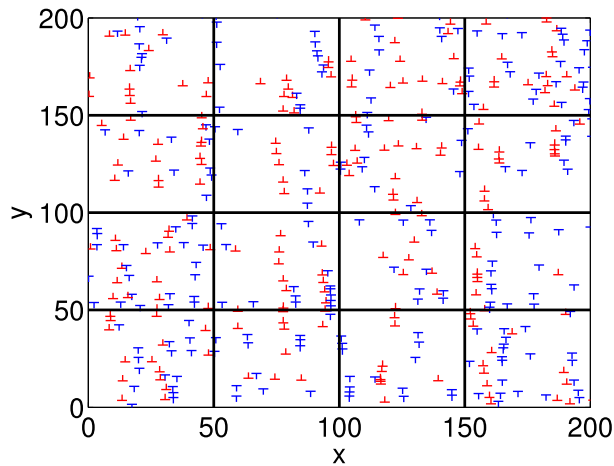}
\caption{A snapshot from the
dislocation system. A two-dimensional system of discrete edge
dislocations subject to an external shear stress $\sigma$ close to
the critical value $\sigma_c \approx 0.025$, showing also the
division into boxes of $l=50b$. The colors (red and blue) indicate
the sign of the Burgers vector of the dislocations. Only glide
motion of dislocations along $x$ direction is considered for
simplicity. We study the spatial fluctuations of the strain rate by
dividing the system into boxes of linear size $l$, and defining the
local strain rates over some time interval $\Delta t$, see Equations
(\ref{eq:local_sr}) and (\ref{eq:local_sr_fluct}).}
\label{fig:dddsystem}
\end{center}
\end{figure}

In the simulations, dimensionless units are used by measuring
lengths in units of $b$, times in units of $1/(\chi D b)$, and
stresses in units of $D$. The stresses are computed by imposing
periodic boundary conditions in both directions. The equations of
motion are integrated numerically with an adaptive step size fifth
order Runge-Kutta algorithm. The system is initially composed of a
random arrangement of $N_0$ such dislocations, with Burgers vectors
${\bf b}=\pm b{\bf u}_x$ parallel to the glide direction (with an equal
number of dislocations with the $+$ and $-$ signs).
To mimic dislocation annihilation occurring in real plastically
deforming crystals, two dislocations with Burgers vectors of
opposite sign are removed from the system if their mutual distance
is less than $2b$. The random initial configurations are first let
to relax in the absence of externally applied stresses. During this
initial relaxation, a significant fraction of the dislocations get
annihilated, leading to a reduction of the dislocation number from
the initial $N_0 = 1600$ for $L=200b$ to the range  of $N=500-600$
after the relaxation. During the subsequent dynamics under the
influence of an external stress $\sigma \approx \sigma_c$, only a
small amount of further dislocation annihilations take place.
Notice that the relaxation process described above before applying
the external stress is essential to obtain the Andrade law:
initial conditions with randomly positioned dislocations lead to a
different, roughly exponential time dependence of the
strain rate \cite{csikorjstat}.

It is known that the DDD model exhibits a jamming or yielding phase
transition, at a critical value $\sigma_c$ of the shear stress
\cite{oma}. Below that the activity eventually stops as at absorbing
state phase transitions typically. Above, in the thermodynamic limit,
a finite deformation rate exists. By applying a constant external
stress, for the setup described above, we find $\sigma_c \approx
0.025$. For an example of a dislocation configuration observed close
to the jamming threshold, see Figure \ref{fig:dddsystem}. Notice
that the dislocations tend to form various metastable structures,
such as dipoles and walls.

\subsection{Experimental setup}

\begin{figure}[ht!]
\begin{center}
\includegraphics[width=6cm,clip]{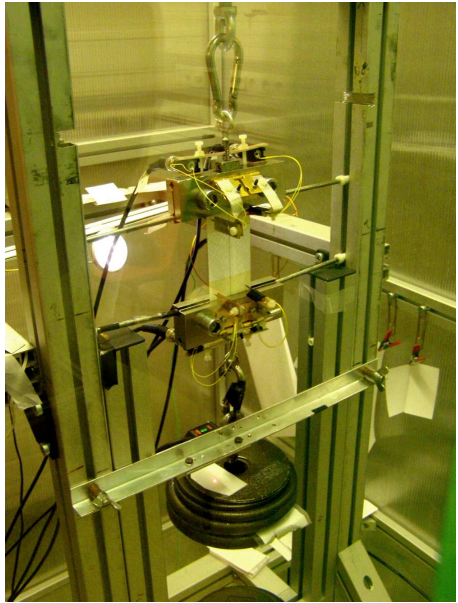}
\caption{The setup for creep experiments.  A load 
is attached to the lower clamp and its movement is controlled by 
using pneumatic cylinders. A camera was attached to the frame, 
and a laser distance sensor followed the movement of the lower clamp. 
Images were taken during the experiment at rate 0.1~Hz...1 Hz.}
\label{fig:experiments}
\end{center}
\end{figure}

\begin{figure}
\begin{center}
\includegraphics[width=10cm,clip]{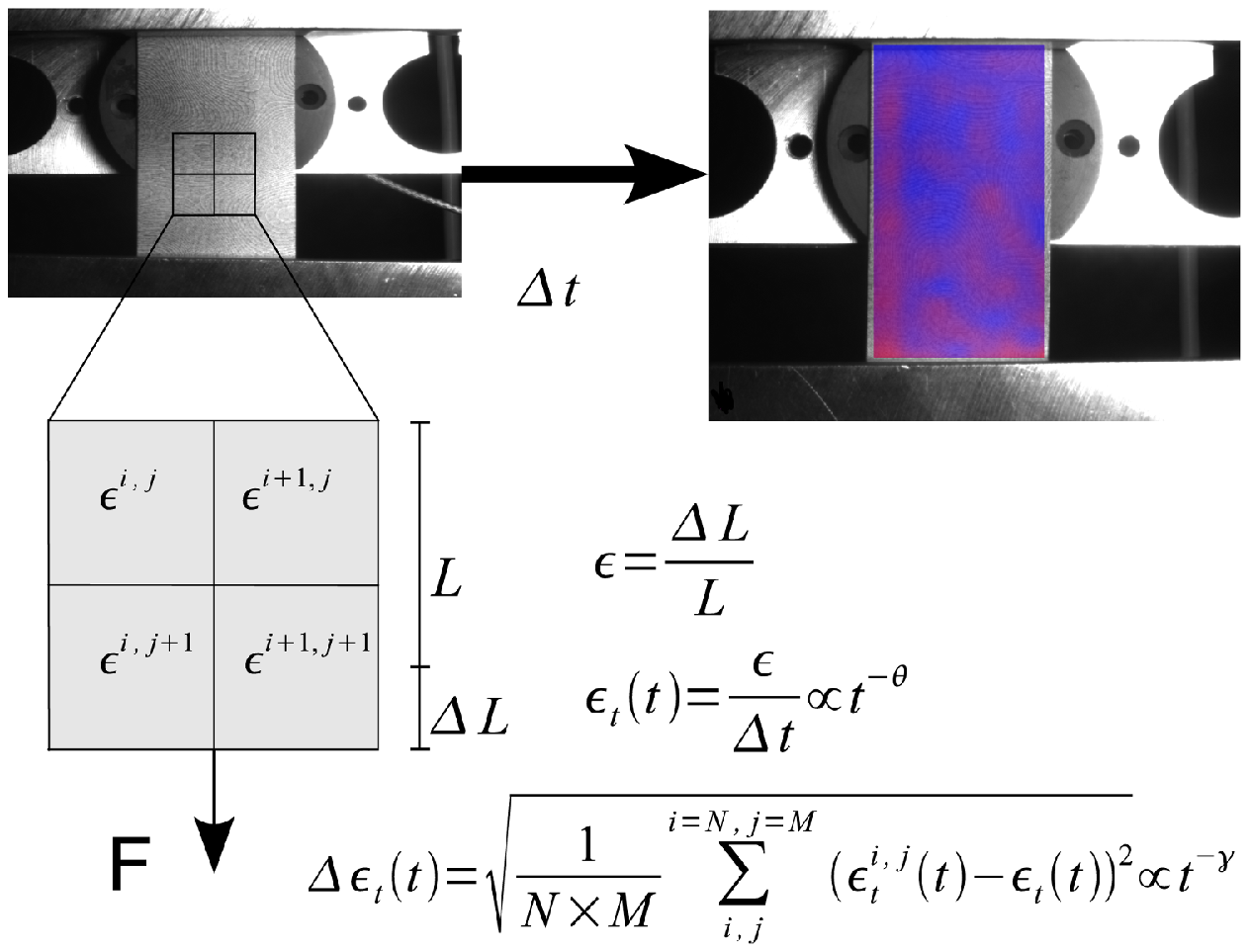}
\caption{The experimental scenario. From a pair of 
digital images at a time-interval $\Delta t$, the local deformations 
are extracted at a grid. The strain fluctuations are measured
via the time-dependent standard deviation, and compared to the mean creep rate,
here in primary/Andrade creep. In the DDD simulations a similar approach is
used, by subdividing the system of linear size $L$ into boxes of linear size $l$ 
in which the local strain rates $\epsilon_t^{i,j}$ are measured. A digital
image of a paper sample on a scale of 40 mm$^2$. Superposed is a typical deformation grid for a
time difference $\Delta t$ of 10 seconds. The color scale indicates the degree
of local creep deformation (blue: small, red: large). In the
background: the experimental setup. The visible speckle pattern has
been printed, and designed to have a structure and contrast
appropriate for the DIC method on the scale at hand. Figure reproduced
from \cite{main}.} \label{fig:scenario}
\end{center}
\end{figure}

The experimental setup is shown in Figures \ref{fig:experiments} and 
\ref{fig:scenario}. The lower clamp is connected to a set of metal 
weights. Typical loads were around 40~N which corresponds to a stress of 
13~MPa. The samples were 100~mm~$\times$~30~mm slices cut from standard 
office paper, loaded in the ``cross-machine" direction.
Relative humidity and temperature were kept in constant conditions at
30\% and 22 C$^{\circ}$, respectively, and the applied load was adjusted to 
achieve different creep times-to-failure, which generally were between 
10 and 60 minutes. In the experiment, the sample is fixed between steel 
clamps mounted on an aluminum frame and a steel jig is used to ensure 
the correct alignment of the sample between the clamps. The sample is 
imaged during the experiment with a digital camera. A laser distance 
sensor is used to measure the elongation of the sample together with 
a digital image correlation analysis.

A set of experiments with a larger magnification was also made. The
typical image size was about 3~mm~$\times$~4~mm and the samples were
specially prepared laboratory sheets, where 5\% of the fibres were
treated with colour prior to fabrication or sheet making to enhance
the contrast and to produce a sharp natural-type pattern. The larger
magnification makes it possible to reduce the noise level, which is
relative to digital image dimension, and to obtain the strain rates
and their fluctuations throughout the entire experiment.

The sample was imaged during the experiment with PCO's 1 mega pixel
grayscale digital camera, SensiCam 370KL0562. The camera has a low
thermal noise ratio due to the cooled CCD and 12 bit grayscale
resolution. The exposure time in measurements was 200 ms. The
sampling frequency of the digital imaging is in the range 0.1...1Hz.

\subsection{Strain field measurement in experiments}

\begin{figure}[t]
\begin{center}
\includegraphics[
width=7cm,clip]{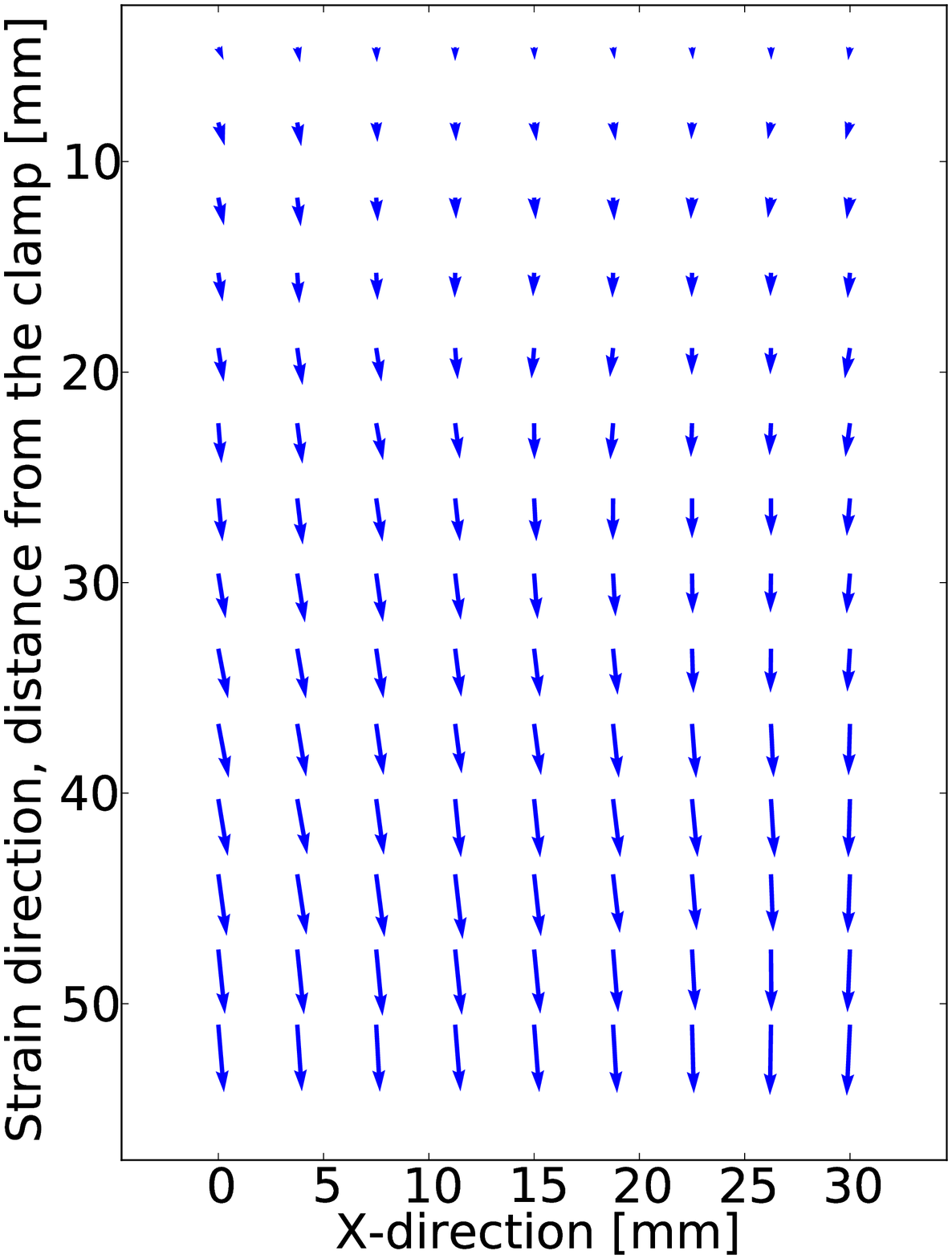}
\caption{Vector field of absolute displacements at $t=10\, s$
in the creep experiment on paper. Vector lengths are scaled so as to
just avoid an overlap and thus show the relative differences
 of displacement vectors. Sample time to failure $t_c =
1549\,s$. }
\label{fig:vecfield}
\end{center}
\end{figure}

The strain is defined on an evenly spaced grid on an image, 
which consists of a discrete set of points $(i,j)$ with constant 
spacing $\Delta d$. $j$ refers to the strain direction. From 
the DIC one obtains displacements on each point $\Delta y^{i,j}$ 
in a time interval $dt$. A spatial strain rate $\epsilon_t^{i,j}$ 
in a grid point $(i,j)$ is computed using
\begin{equation}
\epsilon_t^{i,j} = \frac{\Delta y^{i,j+l/2} - \Delta y^{i,i-l/2}}{l dt}
\label{eq:strain}
\end{equation}
where $l$ is the length scale within which the strain is measured. 
The distribution $P(\epsilon_t^{i,j})$ ensues, which
describes the strain rate evolution during the creep
experiment. An example of the displacement distribution is shown 
in Figure \ref{fig:vecfield}, and the computation of the local strain
rates is illustrated schematically in Figure \ref{fig:scenario}.

The starting point of the DIC algorithm is an elastic image
registration algorithm \cite{kybic,hild,sutton}. The algorithm
describes both the image and the deformation using the B-spline
model. The algorithm finds the deformation function by using the
multiresolution approach for the minimization, so that the image and
the deformation model is refined every time the convergence is
reached. The criteria for the convergence is the sum of squared
differences (SSD), i.e. the deformation function is applied to the
original image and the SSD is computed against the deformed image. The
algorithm is described in \cite{kybic2}. The knots of the
B-splines were defined in an evenly spaced grid, with knot spacing
$h \times h$ pixels. The exact algorithm for the deformation
computation is described in \cite{kybic2}. Knot-spacings, i.e. crates
16x16 to 64x64 pixels were used in the computations. One
method to estimate the error of the displacement field is by using
different crates for same image pairs: the difference of
deformations was computed using two different crates and it was of the order
of 0.003 pixels. As a conclusion of these tests in various experiments
\cite{main, mikapahvi} we can expect that the algorithm is able to
find large continuous deformations with an accuracy of about 0.1~mm in
the sample-scale imaging. The out-of-plane deformations put a limit to
the accuracy, since the paper thickness is the order of 0.1~pixels in
the sample scale experiments.

Note that the sampling frequency
of our imaging is such that we do not expect to see the single,
microscopic "plastic" or yield events \cite{events}. 
That is, our
results do not reflect directly on the microscopic nature of the
dynamics that leads to the fluctuations we actually measure. Thus we
cannot conclude much on the elementary processes and their
interactions in space and time, except in the coarse-grained sense.
This is actually also quite true for the DDD model, though all the
dynamics is carried out by the mobile dislocations - figuring out
the causality of the intermittent dynamics is not easy. One should
also point out that we measure with the technique localized strain
fluctuations, which however are not a priori nor a posteriori
related to any localization in the sense of the formation of
``strain bands'', or ``yield bands''. A separate study should be
done on the possibility of observing related phenomena in the
tertiary creep phase.

When defining and measuring the fluctuations of the strain-field we
study only the $y$-component of the strain. 
This is a simplification since also transverse strains are observed. 
This is seen in Figure~\ref{fig:vecfield} where we show the vector 
field of absolute displacements taken from the primary creep regime 
of a sample whose total time to failure was $t_c = 1549\,$s. As can be
seen, the $y$-component is clearly the dominating component of 
deformation.

It is furthermore important to point out that the local strain
(rate) fluctuations are in no obvious way connected directly to the
disordered structure. Figure \ref{fig:scenario} illustrates also
partly the variations in the sheet transparency, which is related to
the local porosity and to the mass per unit area and its
fluctuations. These are in turn correlated in a non-trivial way with
the local elastic and inelastic material response.

\subsection{Measuring acoustic emission and typical creep curve}

\begin{figure}[t]
\begin{center}
\includegraphics[
width=12cm,clip]{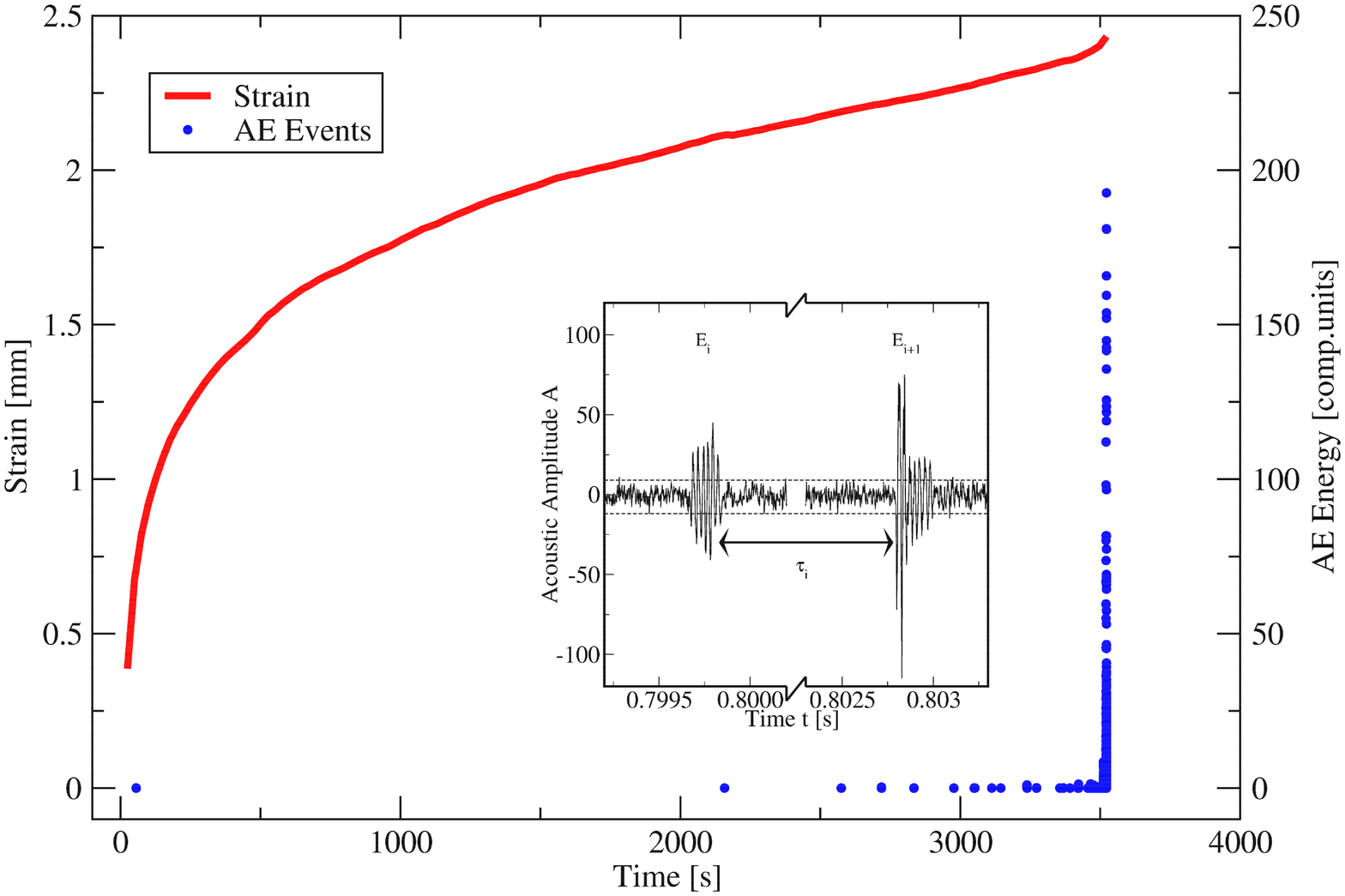}
\caption{Typical time vs. strain curve from a creep experiment, with
also the AE energy shown as a function of time. Red
curve shows the  total strain of the sample and  the blue dots
indicate acoustic emission events during the experiment, with the
energy scale indicated on the right. Inset: zoom to the acoustic
emission signal, where we observe two consecutive events with
energies $E_i$ and $E_{i+1}$ greater than a given threshold and a 
waiting time $\tau_i$. This figure shows typical creep behaviour of 
paper, with the three phases (primary, secondary, tertiary). It 
also indicates the relative absence of crackling noise, that is 
the small number of acoustic emission events (AE) during the primary 
(no events) and secondary creep of the sample.} \label{fig:virakuva}
\end{center}
\end{figure}

The strain from an experiment together with acoustic emission (AE)
events is presented in the main figure of Figure \ref{fig:virakuva}. 
The inset shows an example signal from an
acoustic emission timeseries, with the events shown.

The AE measurement system consists of a piezoelectric transducer, a
rectifying amplifier and works with continuous data-acquisition. The
time-resolution of measurements is 10 $\mu$s and the
data-acquisition is free of dead time. During the experiment we
acquire bi-polar acoustic amplitudes by piezocrystal sensor, as a
function of time. The transducer is attached directly to paper and
no coupling agent is used. Data acquisition channel has 12-bit
resolution and a sampling rate of 312 000/s. The transmission time
from event origin to sensors is of the order of 5 $\mu$s. Acoustic channels
are first amplified and after that saved to a hard disk. The shape
of the AE pulses can change and attenuate during
transmission, but that should have only a minor effect on our
analysis. Post-processing (thresholding) the  amplitude signal we find a discrete
sequence of events with corresponding energies $E_{i}$ and arrival
times $t_{i}$. Figure \ref{fig:virakuva} shows that in a typical test,
one observes the usual creep phases if the test is allowed to
continue until final failure. The important point is that actual AE
starts only close to the final failure of the sample.

\begin{figure}[t]
\begin{center}
\includegraphics[
width=8cm,clip]{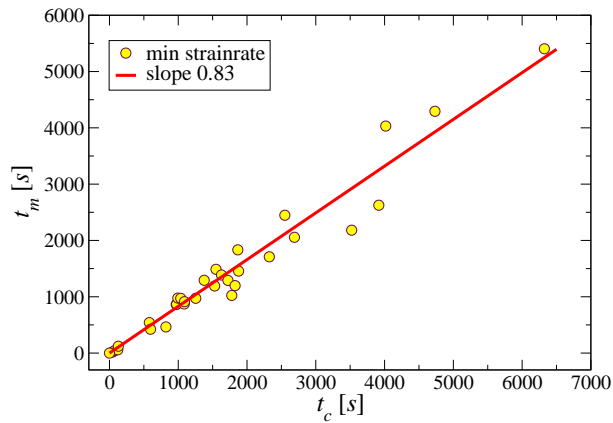} \caption{Time to failure vs. time to minimum
of the strain rate from the experimental data. Time $t_m$ to the
strain rate minimum is proportional to time to failure $t_c$. The
best-fit proportionality constant indicates $t_m \sim 0.83 t_c$.
This is similar to the typical Monkman-Grant relation \cite{monkman}.} \label{fig:tmvstc}
\end{center}
\end{figure}

It is interesting to note that there is a large degree of
variability (Figure \ref{fig:tmvstc}) in the typical timescales of the
creep process from sample to sample, but that the minimum creep rate
time $t_m$ and the failure time $t_c$ are linearly related: a Monkman-Grant
-like universal behavior exists such that $t_m \sim 0.83 t_c$
\cite{monkman, nechad}. In other words, there is a degree of
universality in the behavior of individual samples since the shape
of the creep rate curves is such that the life-time is preceded by
such a minimum around the transition to tertiary creep. The role of
the strain fluctuations for the origins of the $t_m/t_c$-relation
should be further investigated.

\section{Results and discussion}
\label{sec:results}

In this section we review the main results, and discuss briefly their 
interpretation.  At first we consider arguments for the strain
rate fluctuations $\Delta \epsilon_t(t) \sim t^{-\gamma}$. These are
in spirit related to dislocation based theoretical arguments for the
Andrade law presented by Cottrell \cite{cottrell}
and earlier by Mott \cite{mott}. We conclude that as expected these
have problems with the fluctuation scaling, as they do not include
collective phenomena. Then, we link the local creep statistics to
scaling theories of integrated order parameter fluctuations for
absorbing state/depinning phase transitions. This is analyzed both for the
experimental data and the simulations, including a brief study of the
order parameter fluctuations exhibited by the LIM/qEW equation close to
the depinning transition.

\subsection{Strain rate and fluctuations in the DDD model}

In agreement with previous results \cite{miguel2,epjb}, we find that the mean
strain rate for $\sigma \approx \sigma_c$ behaves in the relaxation
phase as
\begin{equation}
\langle \epsilon_t \rangle (t)
= \left \langle \frac{b}{L^2}\sum_{n=1}^{N} s_n b v_n(t) \right\rangle
\sim t^{-\theta},
\end{equation}
with $\theta \approx 2/3$, corresponding to the Andrade/primary
creep law.

\begin{figure}
\begin{center}
\includegraphics[
width=8cm,type=eps,ext=.eps,read=.eps,clip]{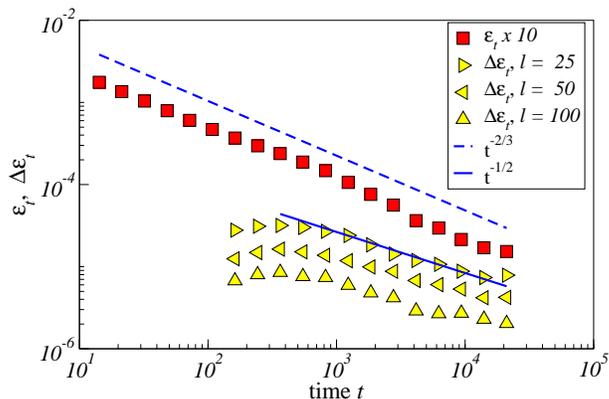}
\caption{The Andrade's law and the
fluctuation scaling for the DDD model for three different box sizes
to compute the local rates, at $\sigma \approx \sigma_c$. The early-time cross-over in the
fluctuations is due to the fixed time-intervals of $\Delta t=100$ at
which they are computed. Figure reproduced from \cite{main}.} \label{fig2}
\end{center}
\end{figure}

We then study the spatial fluctuations of the strain rate by
dividing the system (see Figure \ref{fig:dddsystem})
into boxes of linear size $l$,
and defining the local strain rates over some time interval $\Delta
t$ by
\begin{equation}
\label{eq:local_sr}
\epsilon_t^{i,j} (t) = \frac{b}{l^2} \sum_{{\bf r}_n \in {\mathrm box}~i,j}
s_n b \left[ \frac{x_n(t+0.5\Delta t)-x_n(t-0.5 \Delta t)}{\Delta t}  \right].
\end{equation}
The possible choices for the linear box size $l$ are limited by
$1/\sqrt{\rho}<l<L$, where $\rho$ is the dislocation density (and
thus $1/\sqrt{\rho}$ is the mean dislocation-dislocation distance).
Therefore, for $L=200 b$, we consider $l=25 b$, $50 b$ and $100 b$.
There are practical limits also for $\Delta t$ values that can be
used: these should be much smaller than  the numerically feasible
simulation times (of the order of $10^4 - 10^5$ time units), but
also as long as possible to be able to meaningfully compare the
results with experiments in which two consecutive images are
separated by a ``macroscopic'' time. Therefore we chose to use
$\Delta t = 100$ in the simulations as is visible in Figure \ref{fig2}.

The spatial strain rate fluctuations are quantified by the time
dependence of the standard deviation of the local strain rates,
\begin{equation}
\label{eq:local_sr_fluct}
\Delta \epsilon_t (t) =
\sqrt{\left(\frac{l}{L}\right)^2\sum_{i,j}
(\epsilon_t^{i,j} (t) - \langle\epsilon_t^{i,j}(t)\rangle)^2 }.
\end{equation}
For times $t>\Delta t$, $\Delta \epsilon_t (t)$ is found to
decay like a power law in time, $\Delta \epsilon_t \sim t^{-\gamma}$,
with $\gamma \approx 0.5$. Thus, the magnitude of the strain rate
fluctuations decays more slowly in time than the mean strain rate,
implying that the role of the fluctuations becomes increasingly important
with time.

\subsection{Strain rate and fluctuations in the paper experiments}

Similarly to the results of the DDD simulations, the experimental 
samples are found to exhibit a primary creep
regime characterized by a power law decay of the average creep
strain rate,
\begin{equation}
\langle \epsilon_t \rangle \sim t^{-0.7\pm0.1}.
\label{eq:puolikas}
\end{equation}
The spatial variations are found to be characterized by a power
law time dependence of the standard deviation of the local strain
rates,
\begin{equation}
\Delta \epsilon_t = \sqrt{\frac{1}{N}\sum_{i,j} 
(\epsilon_t^{i,j} (t) - \langle \epsilon_t^{i,j}(t)\rangle)^2 } \sim t^{-\gamma},
\label{eq:puolikas_flucts}
\end{equation}
where the fluctuation exponent $\gamma = 0.55~\pm~0.1$ is found. 

\begin{figure}
\begin{center}
\includegraphics[
width=8cm,type=eps,ext=.eps,read=.eps,clip]{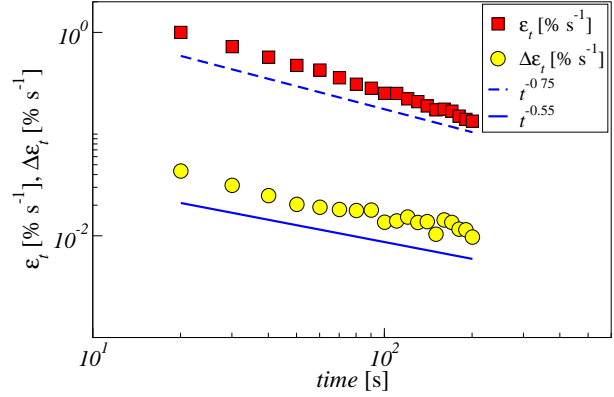}
\caption{Average creep rate and the standard deviation of the local
creep rates from the sample scale experiments. Data for a typical set of 
experiments, $l$ = 60mm and crate 48 $\times$ 48, and SSD = 0.1~pixels. The samples
average lifetime is 800-1600 seconds. The Andrade-to-logarithmic creep
transition appears to take place continuously at 150-200 seconds.
The data is averaged over 16 experiments. The effective Andrade
exponent is about 0.7, in the range 10 to 50 seconds, and becomes
close to one (signalling the onset of logarithmic creep) above 150
seconds. Figure reproduced from \cite{main}.} \label{fig:fsl}
\end{center}
\end{figure}

\begin{figure}
\begin{center}
\includegraphics[
width=8cm,type=eps,ext=.eps,read=.eps,clip]{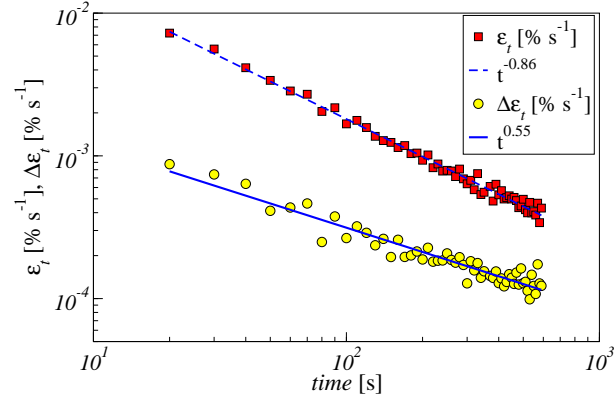}
\caption{An example of the average creep rate and the
standard deviation of a single sample. The crate is 16x16 pixels, 
SSD=0.05~pixels and l=3mm. \label{fig:fslsingle} }
\end{center}
\end{figure}

\begin{figure}
\begin{center}
\includegraphics[
width=8cm,type=eps,ext=.eps,read=.eps,clip]{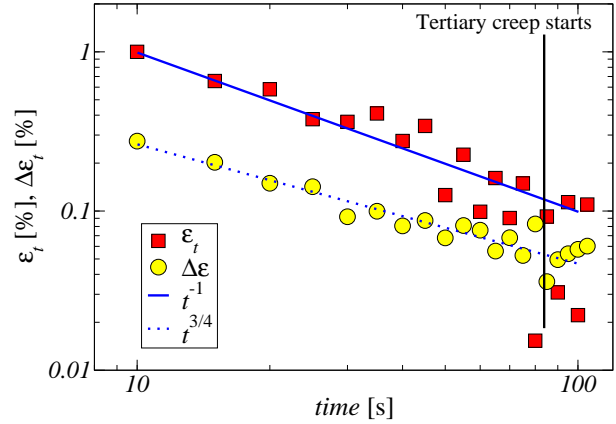}
\caption{Experimental results from logarithmic creep using a smaller
imaging area, 5 second image intervals, $l$=0.3mm, and a smaller load, 
showing the average creep rate and the fluctuations. The former decays 
faster,  and at the end these two become comparable, on the observation 
scale. The data is an average over 9 experiments. Figure reproduced from
\cite{main}.} \label{fig3}
\end{center}
\end{figure}

The magnitude of the fluctuations of the strain rate decay slower with
time than the average creep rate, implying that fluctuations become
more important in relative terms with time. An average over 16 samples is
presented in Figure \ref{fig:fsl}, and a single example is shown in Figure
\ref{fig:fslsingle}. The fluctuation scaling result was independent of
the length $l$ used. In the sample scale images during creep, the
fluctuation scaling was observed using $l=14...60~mm$. Here, the crate is
the lower and the sample size the upper bound of $l$.
At lower stresses, logarithmic creep is found with no
signature of the Andrade phase. The DIC analysis on the
magnified images shows that the relative strength of the fluctuations
increase also during the logarithmic creep phase, 
as evidenced by Figure \ref{fig3}, where $l=0.3$ mm was used.

\begin{figure}[h!]
\begin{center}
\includegraphics[
width=8cm,type=eps,ext=.eps,read=.eps,clip]{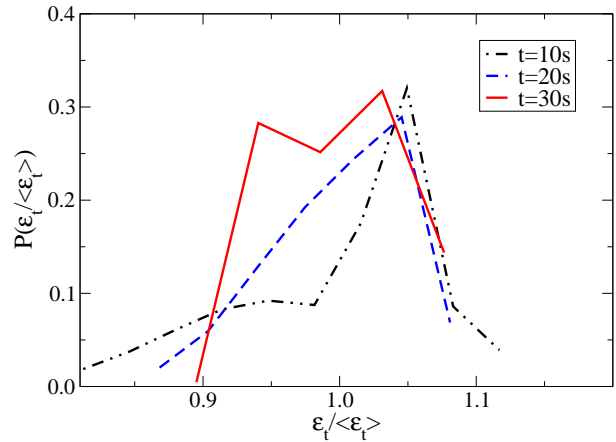}
\caption{The scaled probability distribution functions of strain
rates $\epsilon_t^{i,j}$ from an experiment, where the time to 
failure is 1038~s. The width of the distributions increases when 
time passes. The skewness decreases in this particular sample, but 
the feature is not observed in all experiments.} \label{fig:mikkoscaling}
\end{center}
\end{figure}

The average creep rate and the fluctuation amplitude change with
time so the creep rate probability distribution (PDF) might evolve
as well. Typical examples of the distributions of (the y-component
of) relative strain rates are depicted if Figure \ref{fig:mikkoscaling},
showing that such PDF's become narrower but can be otherwise roughly 
collapsed with the expected exponent.

\subsection{Local stress-fields and elastic modulus fluctuations of paper
samples}

The structure of paper exhibits disorder in many scales
\cite{niskanen}. In this work we have studied fluctuations of the
strain field during the creep experiment which corresponds a
floc-scale and a fiber-scale. The floc-scale corresponds density
fluctuations of paper sheet at scales order of 1~cm. The fiber scale
emerges from 0.1~mm thick fibers and fiber-fiber bonds. Density
fluctuations lead to fluctuations in the local elastic modulus. When
one imposes a constant stress during the beginning  of the creep
experiment, elastic modulus fluctuations lead to an inhomogenous
stress field. The question is can we understand fluctuations in the
displacement/strain rate field based on structural disorder on the
floc- or fiber-scale?

Andrade's law can be expressed in terms of the global strain
$\epsilon \sim t^{1/3}$ or locally $\epsilon_t^{i,j}(t) \sim
t^{-2/3}$. The same applies to fluctuations: The fluctuation scaling law
$\Delta\epsilon_t(t) \sim t^{-\gamma}$ can be studied at different
scales, by changing the size of the region where the local strain
rate is measured, and related to Andrade's law, i.e. to the average strain
rate over the whole region. We have observed universal behaviour
from fiber to sheet scale, suggesting that the data needs a model
which goes below our observation scale; we do not have a similar
microscopic description for paper creep as we have for crystallized
solids based on the concept of dislocation dynamics, i.e. localized
topological defects that exhibit then collective behavior.

The inhomogeneous stress field and its relation to the strain rate
field is related to spatiotemporal aspects of local strain rates
during the experiment: when sample creeps locally it relaxes the
local strain and stress and one would expect that the relative
strain field is not stationary in time. This is seen qualitatively
in accelerated videos of the local creep rates: during the Andrade 
and logarithmic creep, strain fluctuations are not localized into fixed 
regions or zones.

\subsection{Primary creep fluctuations - a simple model}

Here we consider a crude model for such intermittent phenomena. We 
start with the idea of dividing the volume elements/boxes
into two categories according to their instantaneous activity over
the observation time $\Delta t$, i.e. into ``active'' and
``inactive'' boxes. The active boxes are assumed to be characterized
by a local strain rate $\epsilon_t^{a}$ while inactive ones have a
strain rate of zero. The probability distribution of the local
strain rates can then be written as
\begin{equation}
\label{eq:epsdist}
P(\epsilon_t^{i,j}) = p\delta(\epsilon_t^{i,j} - \epsilon_t^{a})
+ (1-p)\delta(\epsilon_t^{i,j}),
\end{equation}
where $p$ is the probability that a randomly chosen box is active.
Next we assume a power law time dependence for both $p(t)$ and
$\epsilon^{a}_t(t)$,
\begin{equation}
\label{eq:peps}
p(t) \sim t^{-\alpha}, \quad \epsilon_t^{a}(t) \sim t^{\alpha-\theta}
\end{equation}
such that the mean strain rate obeys Andrade's law, $\langle
\epsilon_t \rangle = \langle n \epsilon_t^{a} \rangle = \langle n
\rangle \langle \epsilon_t^{a} \rangle = Np \epsilon_t^{a} \sim
t^{-\theta}$. The assumption of two power law -scalings is simply
due to the fact that otherwise it is difficult to see how to obtain
a result similar to the empirical fluctuation results.

The standard deviation of the local strain rates is then given by
\begin{equation}
\label{eq:box_fluct}
\Delta \epsilon_t^{i,j} (t) = (p(t))^{1/2} \epsilon_t^{a}(t) \sqrt{1-p(t)}
\sim t^{-(\theta-\alpha/2)}.
\end{equation}
Thus, to get $\Delta \epsilon_t^i (t) \sim t^{-\gamma}$ with $\gamma
\approx 0.5$, one needs $\alpha \approx 1/3$. This would also imply
$p \sim \epsilon_t^{i,a} (t) \sim t^{-1/3}$, i.e. the number of the
active boxes and the activity within such boxes should have the same
scaling with time. Clearly this kind of scaling argument would need
to be accompanied with a reasoning explaining why the instantaneous
activity should be proportional to the fraction of active
regions/volume elements. One shortcoming of the above crude model is
also that dividing different regions into active and inactive ones
cannot be done unambiguously in systems such as the present ones in
which the local activity is a (broadly distributed) continuous
variable. This makes comparison with simulation results with
the model difficult, as one needs to threshold the local strain
rates in order to subdivide the boxes into active and inactive ones.
Thus, our attempts to verify the above simple model in the simulations
of the DDD model or in the paper experiments were not conclusive.

\subsection{Classical dislocation-based theories of Andrade creep}

Next we will briefly discuss two classical dislocation-based theories
constructed to explain the Andrade power law, and demonstrate how such
ideas could in principle also account for the scaling of the local
strain rate fluctuations. Notice however, that such classical ideas
rely on assumptions that make it difficult to apply them directly to
the systems of the present study.

We start by an argument proposed in a study by Cottrell
\cite{cottrell}. There, the strain rate is written as a product of
the average strain produced by an avalanche, and the number of
``cages'' in the system that initiate such avalanches. By assuming a
linear work hardening law, these two quantities were argued to be
proportional, leading to the Andrade creep law. One possibility to
account for the observed fluctuation scaling would be then to relate
these two quantities to $\epsilon_t^a(t)$ and $p(t)$ in Equations
(\ref{eq:epsdist}), (\ref{eq:peps}) and (\ref{eq:box_fluct}),
respectively, such that scaling of the form of $\Delta
\epsilon_t^{i,j} (t) \sim t^{-1/2}$ would follow from Equation
(\ref{eq:box_fluct}). 
Notice, however, that in the theory of Cottrell
the avalanches are assumed to be initiated by thermal activation,
such that an avalanche is triggered when a dislocation segment
overcomes the pinning force due to a forest dislocation. However,
neither thermal fluctuations nor forest dislocations are included in
the present DDD model, and of course in the paper experiment one does not have
dislocations as such. The underlying implication of the similarity
of a dislocation model is that the Andrade law and the fluctuation
scaling result in general from localized creep events, which then
interact via long-range forces.

Parallel ideas have been earlier presented by Mott \cite{mott}: There,
a system of dislocation sources with the assumption that their
activation is due to incoherent stress fluctuations by the
dislocations that move and thus contribute to the strain was considered.
One can apply the activation idea directly
to individual dislocations, as in the present dislocation system
(DDD). In this case one would take the dislocations to undergo an
intermittent burst giving rise to some characteristic strain
increment once the local stress exceeds some critical value.
Together
with the assumption of a linear hardening law, such an argument was demonstrated to
lead to the Andrade law.
The origin of the linear hardening could then be one of the following: At the critical
point $\sigma=\sigma_c$ of the jamming transition, one could argue
that each dislocation moves/relaxes (on the average) only once
during the experiment, thus leading to a reduction of the number of
potentially moving dislocations as time goes on. This corresponds in
a way to an effective ``hardening'' of the system and gives rise to
the time dependence of the strain rate. Another more convincing
possibility is to assume that the critical values of the local
stresses to initiate dislocation motion depend on strain. To the
lowest order, such a dependence would be linear, corresponding to
the linear hardening law employed by Cottrell and Mott. Again, in this limit,
it could be possible to account for the fluctuation scaling as in the
case of the argument by Cottrell above, by applying ideas presented in
Equations (\ref{eq:epsdist}), (\ref{eq:peps}) and (\ref{eq:box_fluct}).

 As a side note, including higher order terms to the hardening law
 would then result in corrections to the leading Andrade
 scaling/power-law. E.g. a quadratic law for the relation of the
 stress increment vs. strain increment in a tensile test, $\Delta
 \sigma \sim \theta_0 \Delta \epsilon + \theta_1 \Delta \epsilon^2$,
 with the second RHS term dominating) would imply $\epsilon(t) \sim
 t^{1/5}$, or an Andrade exponent $\theta=0.8$.

 As mentioned above, the classical arguments rely on assumptions
 (thermal activation etc.) that cannot always be justified in the
 present case. Therefore, we shall in the following present ideas
 based on the general picture of absorbing state phase transitions,
 which we find more appealing for the systems we study here.

 \subsection{Creep fluctuations in an absorbing state phase
 transition picture}

 It has already been suggested that fluctuation phenomena in plastic
 flow of crystalline solids can be described in the theoretical
 framework of elastic interface depinning \cite{moretti}, with an
 anisotropic long-range elastic kernel (scaling in general as $1/k$
 in Fourier space). Thus, that particular model is very close to the
 mean-field theory (infinite or high dimensional) limit of depinning
 models. It however fails to describe primary creep as in the DDD model,
 and thus does not serve here.

 Depinning describes the movement of an elastic manifold or interface
 in the presence of a driving force, and quenched or frozen disorder.
 In the dislocation plasticity framework, the idea would be to
 coarse-grain from discrete dislocations to a continuum field. Then
 the driving force is the external stress, the disorder is the random
 stress field (from various microscopic origins including large-scale
 dislocation arrangements), and the elastic interactions are
 coarse-grained from the dislocation ones. A depinning transition
 separates a frozen phase from one with a non-zero order parameter,
 for dislocation assemblies the strain rate, when a critical external
 stress value is crossed, and in the generalized phase diagram the
 temperature is taken to zero. In this section we discuss scaling
 properties of fluctuations in creep and compare it to what is expected
 of the integrated order parameter at an absorbing state phase
 transition in full generality.

 The relevant scaling behaviour has been demonstrated using an
 interfacial (height equals integrated activity) description of the
 contact process (CP) \cite{dickman}. The contact process exhibits an absorbing
 state phase transition and belongs to the universality class of
 directed percolation. To review its central properties, the CP is
 usually defined in so that each site of the $d$-dimensional hypercubic
 lattice is either vacant or occupied by a particle. Particles are
 created at vacant sites at rate $\lambda \sim n/2d$, where $n$ is
 the number of occupied nearest-neighbors, and are annihilated at
 unit rate, independent of the surrounding configuration. The order
 parameter is the particle density $\rho$ and the state $\rho = 0$
 is absorbing. As $\lambda$ is increased above $\lambda_c$, there is
 a continuous phase transition from the vacuum to an active state.
 One derives an interfacial model by considering the height of the
 site $h_i(t)$ to be the total amount of time that site is occupied
 (or the number of times it has been active). Dickman and Mu{\~n}oz
 conjecture the scaling hypothesis in Equation (\ref{eq:probheight}) for
 the probability density $p(h; t)$ of the height at any lattice site
 at time $t$ \cite{dickman}. The mean height is $\langle h(t)
 \rangle$, and it is expected that
 \begin{equation}
 p(h; t) = \frac{1}{\langle h(t) \rangle} f(h / \langle h(t) \rangle),
 \label{eq:probheight}
 \end{equation}
 where $f$ is a scaling function. The mean height then obeys
 \begin{equation}
 \langle h(t) \rangle \sim t^{1-\theta},
 \label{eq:cpandrade}
 \end{equation}
 which implies that the variance of the height $\Delta h^2$ scales as
 \begin{equation}
 \Delta h^2 \sim t^{2-2\theta}. \label{eq:cpvariance}
 \end{equation}
 It should be noted that this kind of scaling is expected in general
 in models of rough interfaces with the simplest case of two
 independent (spatial and temporal) scaling exponents. Thus many
 usual depinning models mentioned above adhere to it. For such
 models, the early-time roughening exponent $\beta = 1-\theta$
 (measuring the standard deviation of local order parameter variation
 in time, at a fixed point in space) from the above considerations.
 Moreover as in other cases $\beta$ also describes the one-point
 temporal correlation function in general.

 Fluctuations in the creep deformation of paper and in the DDD can be
 tested against the scaling hypothesis in Equations (\ref{eq:probheight})
 and (\ref{eq:cpvariance}). In the experiments on paper we measure
 local relative strains $\epsilon(x, y; t)$ at positions $(x,y)$,
 where local deformations are computed as displacements between the
 initial loaded state and at a time $t$ from the initial loaded
 state. The picture  from the initial loaded state is taken 1
 second after the load is applied. The scaling hypothesis in Equation
 (\ref{eq:probheight}) can be applied by interpreting the local
 deformations $\epsilon(x,y;t)\equiv h(x,y;t)$ as the local heights
 of an interface and thus the probability density $p(\epsilon ; t)$
 is expected to scale as the one for the local heights in Equation
 (\ref{eq:probheight}). Notice that here we consider the integrated
 strain $\epsilon$, and not the strain rate $\epsilon_t$.

 Equation (\ref{eq:cpandrade}) corresponds to Andrade's law in the
 creep experiment and Equation (\ref{eq:cpvariance}) states that, if
 the scaling hypothesis holds, the variance $\Delta \epsilon^2$
 of the local deformation distribution scales as
 \begin{equation}
 \Delta \epsilon^2 \sim t^{2\beta},
 \label{eq:papervariance}
 \end{equation}
 where $2\beta = 2/3 = 2 - 2\theta$ if the corresponding Andrade's
 exponent in the Equation (\ref{eq:cpandrade}) is taken to be 
$\theta = 2/3$.

\begin{figure}[t]
\begin{center}
\includegraphics[width=8cm,clip]{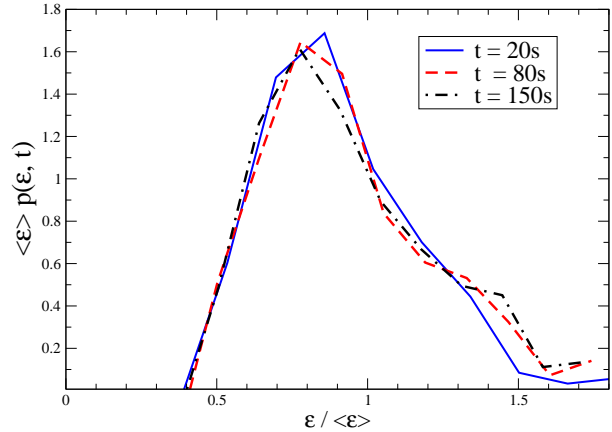}
\caption{Scaling functions of strain fields $f(\epsilon /
\langle \epsilon \rangle; t)$ from experiments on paper 
using $l=14$~mm, crate 16x16 and SSD 0.05 pixels. Different
lines indicate the time at which $f$ is computed, chosen to be at
the beginning, in the middle and at the end of primary creep. The
average distribution over 16 different samples is computed by taking
all the local displacements at a given time and  then computing the
scaling function from the distribution $p(\epsilon; t)$. The scaling
functions collapse and are consistent with the scaling hypothesis of an
absorbing state phase transition (Equation (\ref{eq:probheight})). The
distributions have a similar asymmetric form and a large-strain tail
which is decaying in an exponential fashion.}
\label{fig:epscollapse}
\end{center}
\end{figure}

\begin{figure}[t]
\begin{center}
\includegraphics[
width=8cm,clip]{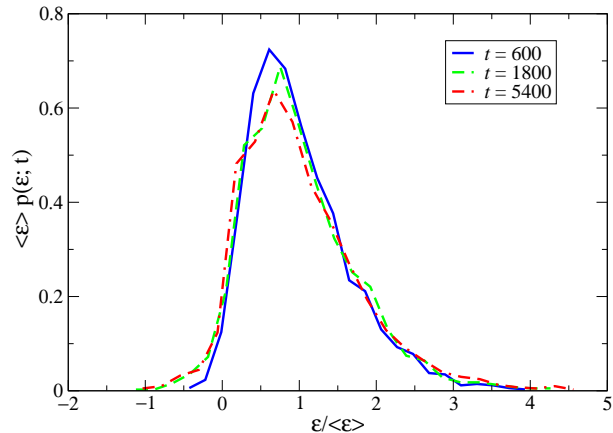}
\caption{Scaling functions $f(\epsilon / \langle \epsilon
\rangle, t)$ from discrete dislocation simulations. Distributions
of the integrated local strain for three different times for the
critical stress value $\sigma_c \approx 0.025$, scaled according to
Equation (\ref{eq:probheight}). The box size used is $l=25b$. For larger
$l$ (not shown) the scaling function becomes more narrow, possibly
approaching the shape of the experimental distributions (Figure
\ref{fig:epscollapse}) for large $l$. The both tails of the
distributions can be well described by an exponential.}
\label{fig:Peps}
\end{center}
\end{figure}

\begin{figure}[t]
\begin{center}
\includegraphics[
width=8cm,clip]{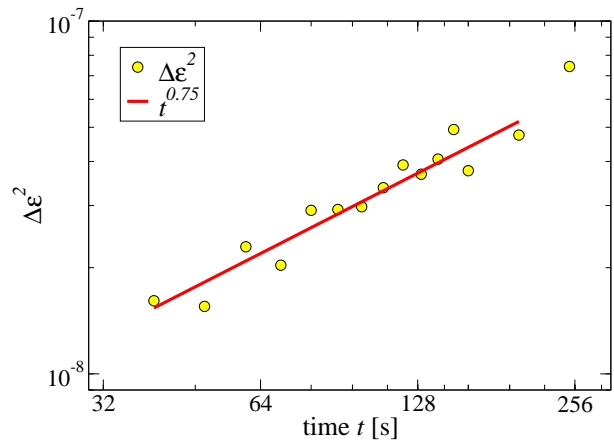}
\caption{Variance of the probability distribution of the total
deformation $p(\epsilon; t)$ as a function of time from experiments
on paper. The result is the averaged variance over 16 experiments.
The scaling behaviour corresponds to the Andrade's scaling of the
experimental data where $\epsilon_t \sim t^{-0.7...0.8}$. This
behaviour is in agreement with the scaling behaviour shown in the
previous figure. } \label{fig:var-peps}
\end{center}
\end{figure}

\begin{figure}[t]
\begin{center}
\includegraphics[
width=8cm,clip]{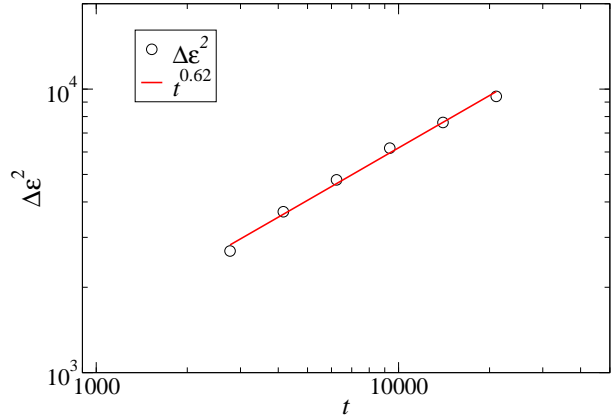}
\caption{Variance of the probability distribution of the total
deformation as a function of simulation time from dislocation simulations,
at $\sigma \approx \sigma_c$.
The data corresponds to that in Figure~\ref{fig:Peps}.
The line indicates a scaling derived from an Andrade law. A least-squares
fit to the data gives a slightly smaller exponent (0.62). The data is 
averaged over 16 samples.} \label{fig:vareps}
\end{center}
\end{figure}

 In Figure \ref{fig:epscollapse} we depict the scaling functions
 $f(\epsilon / \langle \epsilon \rangle; t) = \langle \epsilon
 \rangle p(\epsilon; t)$ during the primary creep in experiments on
 paper. The deformation data $\epsilon(t)$ is averaged over 16
 samples. Similar result for the scaling of the distribution are 
 obtained for the DDD
 simulations as is shown in Figure \ref{fig:Peps}. The variance
 $\Delta \epsilon^2(t)$ is shown in Figures \ref{fig:var-peps} and
 \ref{fig:vareps}. During the primary Andrade
 creep the variance of the fluctuations increases as $\Delta \epsilon^2 
\sim t^{0.75}$ in the paper experiments and close to $t^{2/3}$
 in the DDD simulations. The results indicate that the scaling hypothesis is
 plausible. The slightly larger exponent observed in the case of the paper
 experiment could be related to the experiments being somewhat subcritical.
 One can also extend the experimental data beyond the
 range in time appropriate for Andrade creep, with at least a rough
 agreement with the expected behaviour for the logarithmic creep phase
 (an exponent larger than the $0.75$ for the Andrade regime is
 found). Notice that the result that the spatial fluctuations of the
 integrated strain exhibit the same scaling as their mean is not
 in contradiction with the observation that the fluctuations of the
 {\it strain rate} scale differently from their mean.

 Finally we analyze the spatial correlation functions of the
 integrated order parameter or the creep strain field. 
Given that the experimental strain fields are two-dimensional, we 
can study the correlations in two independent directions. 
The horizontal correlation function is defined as the width
 \begin{equation}
 w_x(r, t) =\sqrt{\langle (\epsilon(x+r, y_c) -  \epsilon(x, y_c))^2\rangle_x},
 \label{eq:corrfunc}
 \end{equation}
 where the roughness of the $y$-directional strain $\epsilon$ is averaged 
 over the $x$-coordinate in a constant position $y_c$. Thus this represents the 
 1-dimensional correlations of the integrated strain for a fixed height
 $y_c$ which we choose to study here for simplicity, given that it is
difficult to fix a reference coordinate system on the sample surface.
Analogously, the vertical correlation function is defined as
 \begin{equation}
 w_y(r, t) =\sqrt{\langle (\epsilon(y+r, x_c) -  \epsilon(y, x_c))^2\rangle_y}.
 \label{eq:corrfunc_v}
 \end{equation}

\begin{figure}[h!]
\begin{center}
\includegraphics[
width=8cm,clip]{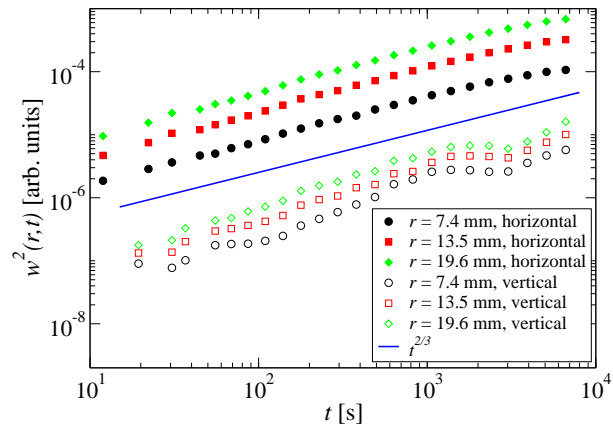} \caption{An example of the squared 
deformation-deformation correlation functions $w(r,t)^2$ (Equation 
(\ref{eq:corrscaling})) from a single experiment with $t_c ~=~ 6324$s,
$l=12.3$ mm, in horizontal ($w_x^2(r,t)$, filled symbols) and vertical ($w_y^2(r,t)$, 
open symbols) directions. The solid line corresponds to the 
$t^{2/3}$ behavior, as derived from the Andrade's law. The statistics of 
the data is insufficient in order to conclude the form of the scaling 
function $f(r,t)$ in Equation (\ref{eq:corrscaling}). Also attempts to
check for the possibility of anisotropic scaling failed due to poor statistic
in the vertical data.}
\label{fig:corrscaling}
\end{center}
\end{figure}

 The scaling behaviour of the width $w(r,t)$ (both $w_x$ and $w_y$) is 
 expected to be governed by the scaling form
 \begin{equation}
 w(r, t) = t^{1-\theta} f\left(\frac{t}{r^{z}}\right),
 \label{eq:corrscaling}
 \end{equation}
 where the above ``simple'' scaling picture has been assumed to hold,
 such that $\beta=1-\theta$. In Figure \ref{fig:corrscaling} we show the 
 squared deformation-deformation correlation functions $w_x^2(r,t)$ and
 $w_y^2(r,t)$ for three different correlation distances $r$ from a typical 
 experiment. For the horizontal direction we observe scaling with an 
 exponent $2-2\theta$ close to 2/3, i.e. consistent with the simple scaling 
 picture and the usual scaling exponent of the Andrade's law, $\theta \approx 2/3$. 
 In the vertical direction the statistics is not so good, and it is 
 not possible to conclude if the system exhibits anisotropic scaling or not.
 The overall relatively poor statistics also made it impossible to test in 
 detail the possible $r$-dependent saturation as implied by Equation 
 (\ref{eq:corrscaling}).

 \subsection{Spatial order parameter fluctuations at the depinning transition}

 For the sake of comparison, we finish by considering the spatial order 
 parameter (=velocity) fluctuations exhibited by driven interfaces close to
 the depinning threshold. To this end, we choose to study the $d=2+1$ Linear
 Interface Model (LIM), or quenched Edwards-Wilkinson (qEW) equation, 
 \begin{equation}
 \frac{\partial h(x,y,t)}{\partial t} = \nu \nabla^2 h(x,y,t) + \eta(x,y,h) + F,
 \end{equation}
 where $h(x,y,t)$ is the local interface height at time $t$, $\nu$ is
 the surface tension, $\eta(x,y,t)$ is a quenched random force term
 with correlations $\langle \eta(x,y,h) \eta(x',y',h')\rangle = 
 2D \delta (x-x')  \delta (y-y') \delta (h-h')$, and $F$ is the
 external force. We simulate the model in continuous time, for linear
 system sizes up to $L=256$. 

\begin{figure}[h!]
\begin{center}
\includegraphics[width=5.25cm,angle=-90,clip]{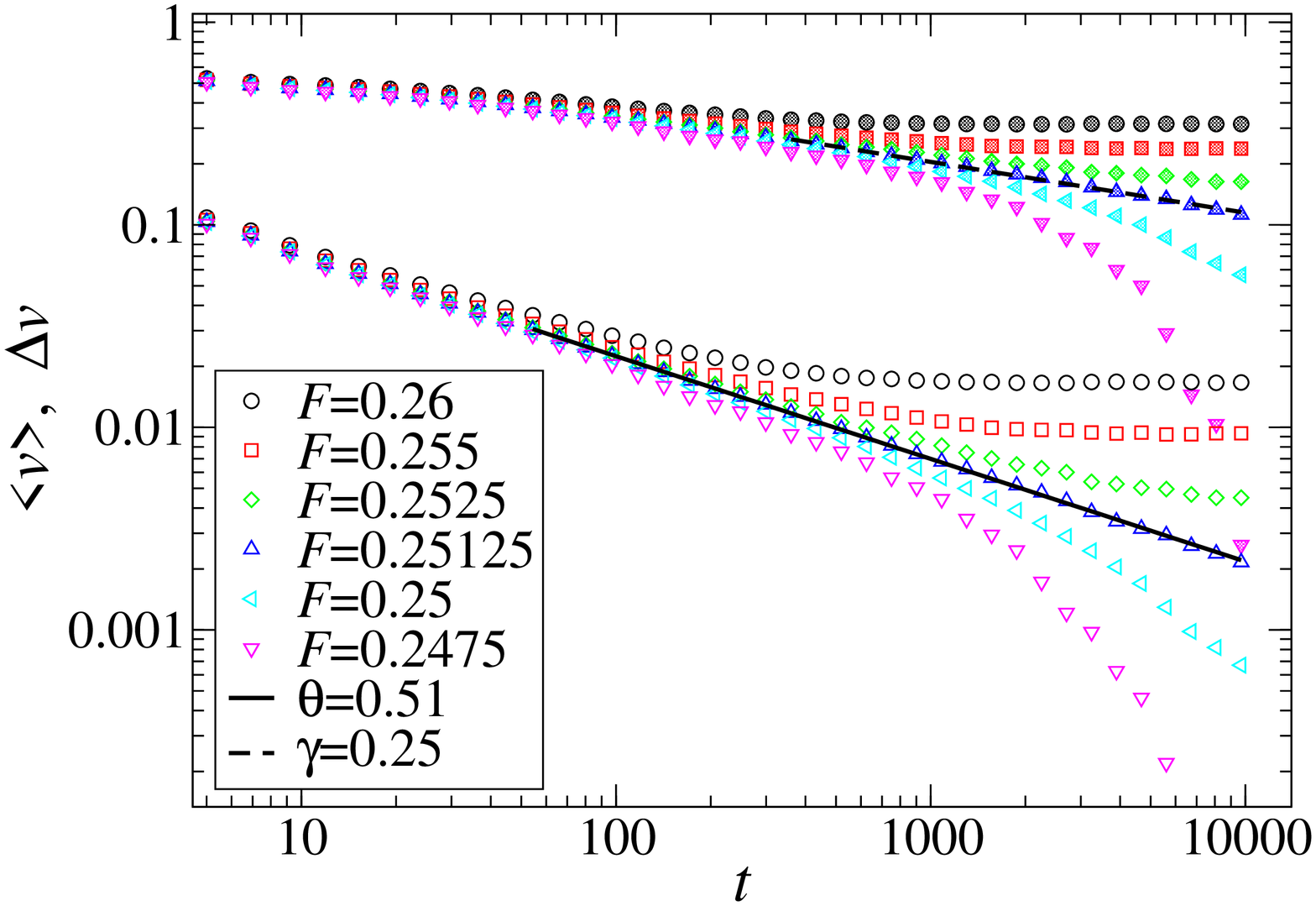}\\ 
\vspace{0.3cm}
\includegraphics[width=5.25cm,angle=-90,clip]{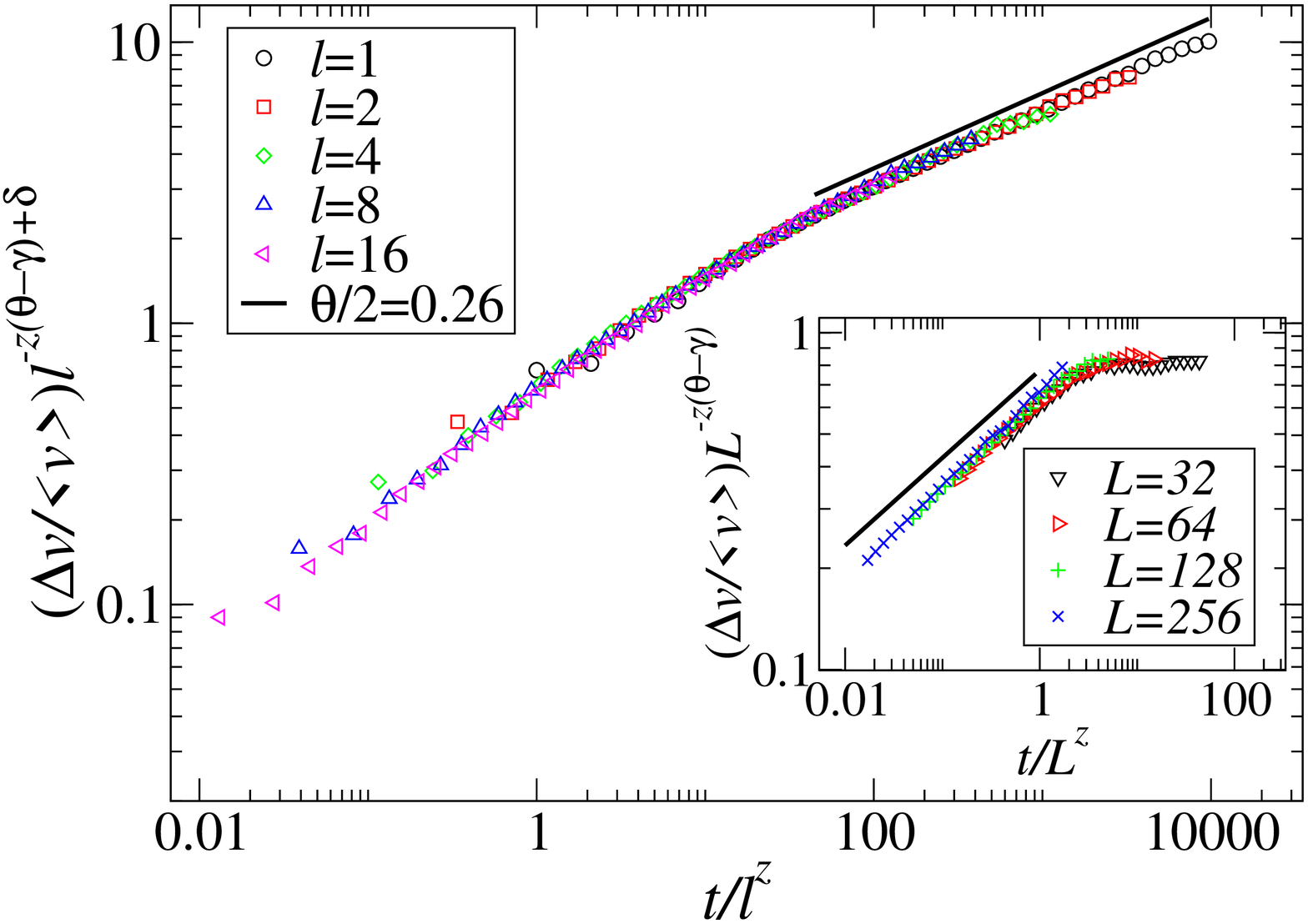}\\
\vspace{0.3cm}
\includegraphics[width=5.25cm,angle=-90,clip]{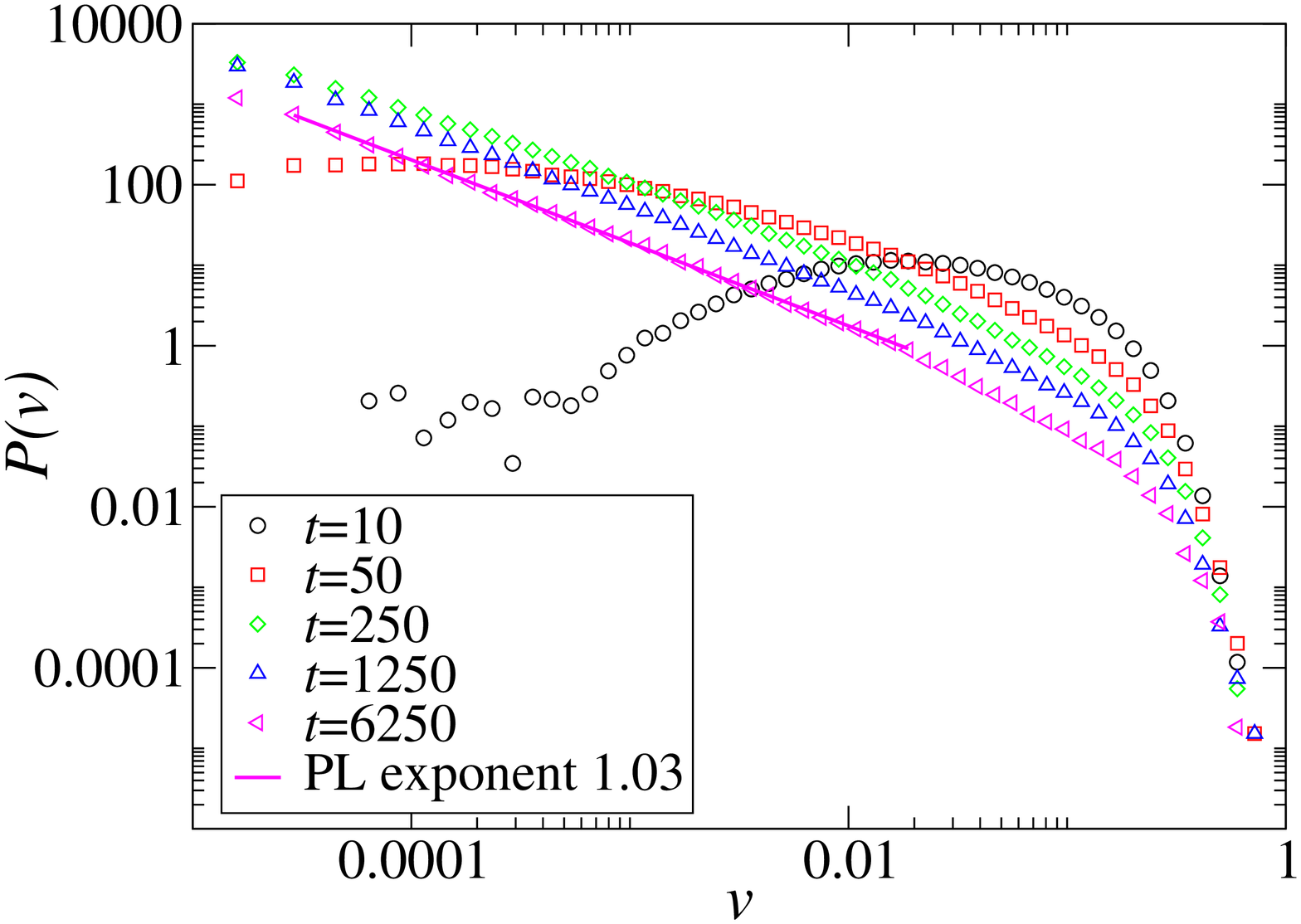}
\caption{Average and standard deviation of the interface velocity
in the $d=2+1$ LIM/qEW. {\bf Top:} Average (open symbols) and standard deviation (filled symbols)
of the local interface velocities for different external forces $F$ in the proximity of
the critical force $F_c \approx 0.25125$, for $l=2$ and $\Delta t=1$. At the critical point, 
both exhibit a power-law decay in time, with exponents $\theta=0.51$ and $\gamma=0.25$, respectively. 
The fluctuation data has been displaced vertically for clarity.
{\bf Middle:} Scaling of the relative interface velocity fluctuations for $F=F_c$ for different 
$l$ (main figure). The inset shows the late time scaling for different 
system sizes $L$. In both cases the data can be collapsed with the dynamical exponent $z=1.56$.
{\bf Bottom:} The distributions of the local interface velocities for different times, for
$F=F_c$, $l=2$ and $\Delta t=1$. The decreasing amplitude in the long time limit implies
that an increasing fraction of the interface is no longer moving.}
\label{fig:lim}
\end{center}
\end{figure}

 The top panel of Figure \ref{fig:lim} shows the spatially averaged
 interface velocity $\langle v \rangle$, and the spatial fluctuations of 
 the local interface velocities $\delta v$ for $l=2$ and $\Delta t=1$ as 
 a function of time $t$ for different values of the external force $F$. 
 Close to a critical value $F=F_c$, both quantities follow asymptotically
 a power law in time, the average velocity obeying $\langle v \rangle
 \sim t^{-\theta}$, with $\theta \approx 0.51$. This is in good agreement
 with earlier results, assuming a scaling relation $\theta=1-\beta$,
 where $\beta \approx 0.48$ is the growth exponent \cite{LES-97}. The fluctuations
 follow $\Delta v \sim t^{-\gamma}$, with $\gamma \approx 0.25$. For
 $F>F_c$ and $F<F_c$, both the mean and the fluctuations deviate from
 the power law, approaching a finite constant value for $F>F_c$, and
 decaying exponentially to zero for $F<F_c$.

 Considering different length scales $l$ and system sizes $L$ reveals
 that the growth of correlations plays a central role in the scaling
 of the fluctuations. The characteristic power law at $F=F_c$ is observed
 only after a transient with an $l$ dependent
 duration. The relative fluctuations $\Delta v/\langle v\rangle$ 
 at $F=F_c$ for different $l$ can be collapsed according to the scaling form
 \begin{equation}
 \label{eq:liml}
 \Delta v/\langle v \rangle = l^{z(\theta-\gamma)-\delta}f(t/l^z),
 \end{equation} 
 involving the dynamic exponent $z\approx 1.56$ \cite{LES-97}, and
 an exponent $\delta$ characterizing the $l$ dependence of the fluctuation
 amplitude. The main figure in the middle panel of Figure \ref{fig:lim} 
 shows a data collapse with $z=1.56$, $\theta-\gamma=0.26$ ($\approx \theta/2$), 
 and $\delta=0.55$. This collapse suggests that the initial 
 transient is related to the time $t^* \sim l^z$ it takes for the 
 correlation length $\xi$ to reach the observation scale $l$. On the
 other hand, the duration of this power law is also limited by the
 finite system size: The late time data for $l=2$ and $\Delta t=1$ 
 collapsed according to
 \begin{equation}
 \label{eq:limL}
 \Delta v/\langle v \rangle = L^{z(\theta-\gamma)}g(t/L^z)
 \end{equation}
 and shown in the inset of the middle 
 panel of Figure \ref{fig:lim} shows that the power law describing
 the relative fluctuations $\Delta v/\langle v \rangle$ extends 
 only up to a time $t^{**} \sim L^z$.

\begin{figure}[h!]
\begin{center}
\includegraphics[width=8cm,clip]{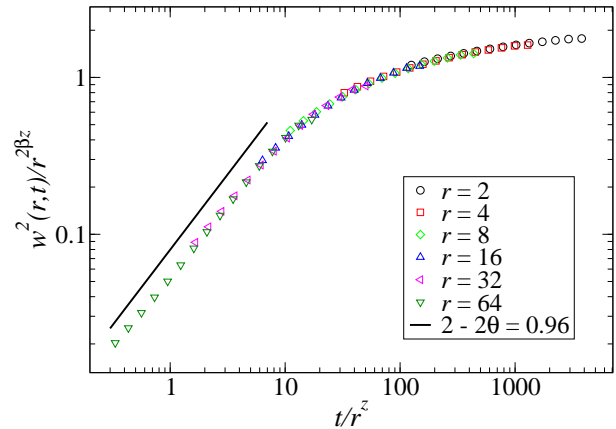}
\caption{Scaling of the squared correlation functions $w^2(r,t)$
in the $d=2+1$ LIM/qEW in the long time limit, for different 
correlation distances $r$. The solid line corresponds to
$2\beta = 2-2\theta = 0.96$, and $z=1.56$ controls the
$r$-dependent saturation.}
\label{fig:lim_correl}
\end{center}
\end{figure}

 To understand the origin of the fluctuation exponent $\gamma$ in this
 case, it is useful to consider the distribution of the local velocities
 (from which $\delta v$ is obtained as the standard deviation) as a
 function of time. The bottom panel of Figure \ref{fig:lim} shows that
 the distribution evolves with time from a peaked one towards a power law with
 a cutoff. By normalizing the distributions by the total number of
 data points (including the ones indistinguishable from zero) reveals
 that the amplitude of the power law decreases with time, but otherwise
 the distribution remains unchanged: Notice in particular how the large-velocity
 cut-off of the distribution is independent of time. Thus, the
 velocity distribution, after the initial transient, can be well
 described by the superposition
 \begin{equation}
 P(v^{i,j}) = pP_m(v^{i,j})+(1-p)\delta(v^{i,j}),
 \end{equation}
 where $P_m$ is the distribution of the mobile interface elements,
 which is found to be independent of time after a transient. The full 
 distribution then evolves only via the decay of $p$, describing the 
 fraction of observation boxes with a finite velocity. Thus,  
 in the case of the LIM/qEW, an argument similar to the simple box argument, 
 Equation (\ref{eq:box_fluct}), leads to a fluctuation exponent $\gamma=\theta/2
 \approx 0.26$, in good agreement with the numerical estimate 
 $\gamma \approx 0.25$.

 Finally we also consider the (isotropic) height-height correlations
 \begin{equation}
 w(r,t) = \sqrt{\langle (h(\vec{r}_0 + \vec{r},t)-h(\vec{r}_0,t))^2 \rangle_{\vec{r}_0}},
 \end{equation}
 analogously to Equations (\ref{eq:corrfunc}) and (\ref{eq:corrfunc_v}).
 After an initial transient, these obey the expected scaling form, Equation
 (\ref{eq:corrscaling}), with the scaling function shown in Figure \ref{fig:lim_correl}.
 The early time growth of the roughness scales with the expected
 exponent $2\beta = 2 - 2\theta = 0.96$, followed by an $r$-dependent
 saturation, controlled by the dynamical exponent $z=1.56$.

 The implications of these results for the creep deformation studies
 are as follows: As the fluctuation exponent $\gamma$ appears to be larger
 than $\theta/2$ in the case of paper experiments and the DDD simulations,
 the box argument, Equation (\ref{eq:box_fluct}), would imply that
 also the strain rate distribution of the active elements of the system 
 would evolve in time, to yield the observed scaling of the fluctuations.
 Detailed analysis of this kind is however difficult as it would
 require a substantial amount of statistics, and is thus out of scope
 of the present study. Analysis involving growing correlations (in the 
 spirit of Equations (\ref{eq:liml}) and (\ref{eq:limL})) could also be complicated
 by the possibly anisotropic nature of such correlations \cite{lasse}.

 \section{Conclusions}
 \label{sec:concl}

 The creep deformation of solids shows, after all, fluctuation
 features that are unexpected. Clearly it is to be expected that in
 any not completely homogenenous and translationally invariant
 material (sample) the rheology has to include in practice a
 fluctuating component. However, the spatiotemporal aspects that it
 then encompasses  first of all are not included in traditional
 materials science descriptions of empirical data, and secondly they
 may - as indeed is the case here - originate from ``collective
 behavior''. And, this is the central point of the current
 investigation.

 The presence of crackling noise in materials is of course an old
 subject, from Barkhausen noise in magnets to acoustic emission in
 the brittle fracture -related deformation of materials and
 geosystems. In these two particular cases, theory has been able to
 explain to a varying degree from almost complete to qualitative the
 features seen such as probability distributions of noise events, and
 correlations in the timeseries measured. An open problem here, and
 in many other similar systems where one can study collective
 phenomena in deformation - such that possess a jamming transition
 for instance - either experimentally or numerically or both, is what
 is the coarse-grained theory like, and what is the origin of the
 eventual criticality and the general picture of the fluctuations
 that ensue. In dislocation plasticity, various attempts of the
 scaling argument -type have been done to understand creep in
 addition to numerical models from DDD to cellular automata. However,
 no formulation exists that would account for the yielding transition
 faithfully. Obviously, in non-crystalline materials the search is
 even further away from a candidate in spite of various
 rheology-based attempts that have been done lately. Thus we conclude
 that experiments such as published here, including the numerical
 ones, should also offer a valuable guideline for future
 developments. 

 In addition to showing that scaling exists in the fluctuations in
 primary (Andrade) and logarithmic creep, we have also shown the
 results of the application of various ideas from absorbing phase
 transitions, including depinning transitions for elastic manifolds
 in random media. It appears that the part of creep that would
 originate from a yielding transition agrees with this picture. Our
 experiments nevertheless suffer from a lack of resolution in space
 and time, and it is to be hoped that a better (two-dimensional) test
 material is found to improve on both. This is true for the creep rate
correlation functions. We note also that the presence of fluctuations
in the logarithmic phase is not a priori easy to relate to a phase
transition picture unlike for the preceding Andrade part. The same
goes also for the untested behavior in tertiary creep as the sample
failure is approached.

 Creep is the simplest paradigm of time-dependent failure. What we
 describe here is the development of creep strain mostly when there
 is not yet actually any damage, i.e. in the classical empirical creep
 language the system is not yet in tertiary creep. This overview raises a wide
 variety of open questions. First, what happens in a ordinary tensile
 test with a certain strain rate, what is the role of the
 viscoplastic deformation in a (apparently) brittle material say, or
 in a very ductile one if one assumes that similar collective
 phenomena are to be found there as well? An analogy from depinning
 is a ramp of the driving force so that eventually the threshold is
 crossed. Second, the role of temperature in depinning (and in DDD
 models) has a specific meaning. It influences i) the mobility of the
 coarse-grained dynamics (scales the timescale in DDD) and ii) it
 leads to "creep" or thermally assisted motion or deformation at the
 end - but beyond the criticality-dependent creep. Third, what is the
 role of such phenomena in materials science, if any? Fourth, one can
 now go beyond creep to various scenarios of time-dependent material
 failure such as fatigue etc., and repeat the experiments and the
 analysis in such cases. One question is the role of relaxation 
 processes or memory effects: how will these influence strain rate
 fluctuations, and if as for elastic manifolds one finds signatures of
 the behavior of glassy systems. An example of such could be aging, 
 for which many separate scenarios could be envisioned.

 In summary, we have discussed in this article by experiments on a
 two-dimensional material (i.e. paper) and by simulating a crystal
 plasticity model, that creep deformation shows fluctuations that can
 be analyzed further using the language of absorbing state phase
 transitions.  For DDD models the existence of the yielding transition
 has been known, and it has been recently established to be a peculiar,
 zero-temperature second-order phase transition. The observations we
 have made are of course related to the critical scaling thereof: how
 exactly is still to be understood. The experimental signatures from a
 non-crystalline material are quite similar, which would hint of a
 greater universality than what one might have expected. Future work
 should include settling some of the open issues listed above, and we
 would also like to underline the possibilities in other systems
 including model ones such as colloidal particle assemblies, with both
 crystalline order and in the amorphous state \cite{video,schall}.

\ack
We thank the support of the Academy of
Finland via the Center of Excellence program and post-doctoral
grants.

\end{document}